**Title**: *Physical Characterization of the December 2017 Outburst of the Centaur 174P/Echeclus*

**Authors**: Theodore Kareta[1], Benjamin Sharkey[1], John Noonan[1], Kat Volk[1], Vishnu Reddy[1], Walter Harris[1], Richard Miles[2]
[1]: Lunar and Planetary Laboratory, University of Arizona
[2]: British Astronomical Association

**Abstract**: The Centaurs are the small solar system bodies intermediate between the active inner solar system Jupiter Family Comets and their inactive progenitors in the trans-Neptunian region. Among the fraction of Centaurs which show comet-like activity, 174P/Echeclus is best known for its massive 2005 outburst in which a large apparently active fragment was ejected above the escape velocity from the primary nucleus. We present visible imaging and near-infrared spectroscopy of Echeclus during the first week after its December 2017 outburst taken at the Faulkes North & South Telescopes and the NASA IRTF, the largest outburst since 2005. The coma was seen to be highly asymmetric. A secondary peak was seen in the near-infrared 2D spectra, which is strongly hinted at in the visible images, moving hyperbolically with respect to the nucleus. The retrieved reflectance spectrum of Echelcus is consistent with the unobscured nucleus but becomes bluer when a wider extraction aperture is used. We find that Echeclus's coma is best explained as dominated by large blue dust grains, which agrees with previous work. We also conducted a high-resolution orbital integration of Echeclus's recent evolution and found no large orbital changes that could drive its modern evolution. We interpret the second peak in the visible and near-infrared datasets as a large cloud of larger-than-dust debris ejected at the time of outburst. If Echeclus is typical of the Centaurs, there may be several debris ejection or fragmentation events per year on other Centaurs that are going unnoticed.

**Corresponding Author:**
Theodore Kareta, tkareta@lpl.arizona.edu
Lunar and Planetary Laboratory, University of Arizona
1629 E. University Blvd., Tucson, AZ, 85719

**Body**:

1.0 – Introduction

The Centaurs are a group of small solar system objects on giant planet-crossing orbits that are sourced from the trans-Neptunian populations and, in turn, are the source of the Jupiter Family Comets (JFCs) in the inner solar system (see a recent review by Dones et al., 2015). Centaurs have dynamical lifetimes in the giant planet region of $10^6 – 10^7$ years, and approximately one third actually end up as comets in the inner solar system (Sarid et al., 2019, see also Tiscareno and Malhotra, 2003, Horner et al., 2004, DiSisto and Brunini 2007). A primary aim of Centaur research is to understand how Centaurs evolve during this time period such that observations of the JFCs can be more directly related back to their outer solar system origins.

(60558) 174P/Echeclus (provisional designation 2000 EC98, hereafter Echeclus) is a Centaur discovered by Spacewatch in 2000 that is among Centaurs displaying comet-like activity. Estimates vary for the fraction of Centaurs that are active, from ~13% of the observed population (Jewitt, 2009) to even lower (Cabral et al., 2019). Echeclus's 60 km diameter (Bauer et al. 2013) is comparable to 29P/Schwassmann-Wachmann (~54 km, Stansberry et al., 2004) but smaller than the largest Centaurs like (10199) Chariklo (~250 km, Bragas-Ribas et al., 2014) or 95P/Chiron (~140 km, Groussin et al., 2004). The orbital parameters and physical properties of Echeclus are listed in **Table 1**. Echeclus's activity is not like traditional insolation driven cometary activity (whereby increased temperatures closer to perihelion drive ices to sublimate more vigorously and drive off more dust). Echeclus has instead been characterized by four very large outbursts of activity in 2005, 2011, 2016, and 2017, each of which increased its visual brightness by several magnitudes with little activity in between. A summary of these four outbursts with sources for further reading is listed in **Table 2**.

Echeclus is best known for the 2005 outburst, in which its visual brightness increased from V~21 to V~14 and a small condensation in the coma was observed to move away from and then back towards the nucleus. This has been interpreted by many to be a fragment or semi-coherent mass of material ejected from the nucleus at the onset of outburst which may have remained active post-ejection (Bauer et al., 2008, Fernandez, 2009). The measured dust production rate (170-400 kg/s) in 2005 was approximately ~30 times higher than that for other Centaurs seen at comparable distances, and the dust particle size frequency distribution was dominated by large particles and similar to "normal" cometary activity as opposed to an impact-related process (Bauer et al., 2008). In later images (March, 2006), no individual fragment was seen, suggesting that it was either small or had disintegrated by then (Rousselot, 2008, Fernandez, 2009). This lack of any identifiable point source in later images has been viewed as both consistent (Fernandez, 2009) and inconsistent (Rousselot, 2008, Rousselot et al., 2016) with a fragmentation scenario. If the secondary source can be attributed to a fragment, the fragment was likely active on its own, small, and perhaps itself falling apart. A non-fragmentary source would require a unique coma process that has not been seen in other comets. The work of Rousselot et al., 2016 did some morphological modeling of the 2005 outburst, but matched their model to images where the condensation was radically less prominent than those seen in Bauer et al., 2008. No further behavior of this type has been observed from Echeclus during any of its subsequent outbursts, nor has it been observed in any other Centaur.

**Table 1 – Orbital Parameters and Physical Properties of Echeclus**

| a [AU] | e | i [deg.] | Period [years] | Diameter [km] | Albedo |
|---|---|---|---|---|---|
| 10.69 | 0.456 | 4.34 | 34.96 | 59 +/- 4 | 0.077 +/- 0.015 |

Orbital data from JPL Horizons for the epoch April 28.0, 2019 (JPL Orbit Solution 90). Physical Size and albedo from Bauer et al., 2013.

**Table 2 – List and Properties of Echeclus's Outbursts**

| Outburst Date | Visual Magnitude Brightening | Notable Features | Relevant Sources |
|---|---|---|---|
| 2005 (December) – 2006 (Late Spring) | ~7 (21->14) | 'Normal' Cometary Activity, Moving 'fragment'. | Choi et al., 2006a, Weissman et al. 2006, Bauer et al., 2008. |
| 2011 (May) | ~2-3 | A 'jet-like feature'? | Jaeger et al., 2011. |
| 2016 (August) | 2.5-3.0 | None reported. | Miles et al., 2016. |
| 2017 (December) | ~4-4.5 | None reported. | James, 2018 |

The processes by which Centaurs become active is an area of active study. Most active Centaurs seem to have recently had orbital changes (Fernandez et al., 2018) which would induce changes to their thermal state. While the Centaurs are at heliocentric distances too large for water-ice sublimation to be a primary driver of activity, the orbital distribution of active Centaurs is broadly consistent with activity driven by the exothermic crystallization of amorphous water ice (Jewitt, 2009, Guilbert-Lepoutre, 2012). Another possible driver of Centaur activity is the exothermic dissolution of trapped gases in the interior (Miles 2016b). Information on recent orbital changes obtained from high temporal resolution orbital integrations can be critical to constraining why a particular Centaur is or is not active, and what substances or processes might be driving it.

Carbon Monoxide (CO) has also been detected in three active Centaurs, 29P/Schwassmann-Wachmann (hereafter 29P, Senay and Jewitt, 1994, Crovisier et al., 1995, Womack, Sarid, and Wierzchos, PASP 129, 2017), 95P/Chiron (Womack and Stern, 1997), and 174P/Echeclus (Wierzschos, Womack, and Sarid, 2017). The detections of CO emission towards both Chiron and Echeclus were weak and should be weighed carefully until confirmed. The detection of CO emission in Echeclus (as well as 29P, Senay and Jewitt, 1994, Crovisier et al., 1995) was found to be slightly blue-shifted compared to the nucleus (Wierzschos, Womack, and Sarid, 2017), suggesting emission from material moving towards the observer from the sunlit side and thus CO ice in thermal contact with the surface. Similar to Echeclus, 29P's activity is dominated by explosive large outbursts, except that at 29P they are periodic in time (Trigo-Rodriguez et al., 2010, Miles, 2016a), a property unique to 29P among known active comets and Centaurs. However, 29P, unlike Echeclus, is also active in between its periodic outbursts (Womack, Sarid, and Wierzchos, 2017).

The surface properties of Centaurs fall into two groups (e.g., Peixinho et al., 2003, Tegler et al. 2008, see also Tegler and Romanishin, 1998) based on their reflectance spectra: the 'more red' group and the 'less red' group, a dichotomy also shared by objects in their trans-Neptunian source populations (e.g., Peixinho et al., 2012, see Jewitt, 2018, Marsset et al., 2019). The 'less red' objects are spectrally neutral or slightly red, while the 'more red' objects have large positive slopes. The active Centaurs, including Echeclus (Guilbert et al., 2009), are primarily in the 'less red' group, suggesting an evolutionary trend whereby the very red surfaces of some Kuiper Belt Objects are apparently systematically removed as Centaurs evolve both in activity and heliocentric distance. The surface of Echeclus has no ice absorption features (Guilbert et al., 2009, Seccull et al., 2019) and is more steeply 'red' at visible wavelengths than near-infrared ones, as is typical for the less red group. In terms of reflective properties and albedo, Echeclus is representative of a 'typical' active Centaur. Sample sizes, while growing, remain smaller than needed to tease out and understand differences between the active and inactive Centaurs not based on their reflective properties (Cabral et al., 2019, see also Bauer et al, 2013). However, the population remains biased but sample sizes remain small due to the low albedos (Bauer et al., 2013) and large heliocentric distances of the objects involved.

The December 2017 outburst of Echeclus was the strongest since the 2005 event and is the subject of this paper. We present visible-wavelength imaging (Sloan r' filter) and near-infrared spectroscopy (0.7 – 2.5 microns) taken in the first week after the onset of Echeclus's outburst. In Section 2, we outline our observations. In Section 3, we describe our data and make preliminary comparisons to previous observations. In Section 4, we present the results of a high-resolution orbital integration of Echeclus's recent orbital history. In Section 5, we present an integrated discussion of the properties of Echeclus's dust and the evidence for a new condensation in the coma of Echeclus related to this outburst, and how Echeclus compares to other Centaurs.

2.0 – Observations and Data Reduction
Below we describe visible wavelength imaging in the Sloan r' filter taken on December 9th, 12th, and 13th of December, 2017 followed by near-infrared spectroscopy from 0.7-2.5 microns taken on the 13th of December, 2017.

2.1 – Visible-Wavelength Imaging
The December 2017 outburst was first reported online by Brian Skiff (Lowell Observatory)[1] on December 7th, UTC. In response, a series of visible-wavelength images in the Sloan r' filter were obtained with the Faulkes North (2 meter aperture, Haleakala, Hawai'i) and South (2 meter aperture, Siding Spring, Australia) telescopes remotely in the hours and days following the report of the outburst. Observational details are listed in **Table 3**. The images were reduced in the way typical for CCDs whereby bias and dark-count frames were subtracted from the science image followed by a flat-field correction. Sky subtraction was completed later during the analysis process. Seeing was ~1.1" – 1.3" on all nights.

2.2 – Near-Infrared Spectroscopy
Near-infrared spectral observations were undertaken during an already-in-progress observing run at the NASA IRTF (3 meter aperture, Mauna Kea). The SpeX (Rayner et al., 2003) instrument was utilized in the low-resolution (R~200) 'Prism' mode with a 0.8" wide slit. Observations of the science targets were 'bookended' by observations of a local G-type star before and after for full telluric correction, and a proper well-studied Solar Analog star SAO 93936 was observed for further correction related to the differences between the local G-type and the solar spectrum as approximated by the Solar Analog star. All observations were conducted at low airmass (AM < 2.0) and at or within several degrees of the parallactic angle, the position angle on the sky at which atmospheric diffraction has the smallest effect on the retrieved spectrum. Reduction was completed primarily within the 'spextool' (Cushing et al., 2004) set of codes written in IDL as well as custom-written scripts in Python. While seeing was quite good (~0.5-0.7"), guiding on target was challenging due to the very low surface brightness of Echeclus in the NIR as described further in Section 3. A more aggressive guiding routine (longer exposures on the guider camera as well as sky-subtraction) was adopted in the second block of observations resulting in much higher SNR (~2-3 times higher depending on wavelength) and tighter PSFs. No observations showed evidence for persistence on the detector, and only data where Echeclus was fully in the slit throughout the entire exposure were used in any analysis.

---

[1] https://groups.yahoo.com/neo/groups/mpml/conversations/messages/33519

Spatial cuts along the second block of NIR spectra (e.g., profiles along the spatial axis of the extracted 2D spectra) are shown compared with line-cuts across the visible-wavelength images at the same position angle in **Figure 1**. The NIR spatial profile shows two peaks which are seen to track together at the same non-sidereal rate for all used second block spectra. The visible linecuts show the asymmetry seen in the images (**Section 3**), but a slight bump can be seen in the December 13th data. More conclusively, when the H-band profile is blurred down to the resolution of the visible data, the two appear quite similar. The secondary peak is 2.45" – 2.60" away from the primary peak. This secondary peak, seen directly in our NIR data and suggested strongly by our visible data, is discussed in **Section 5**.

**Table 3 – Description of Visible and Near-Infrared Observations**

| Telescope | Target | Date | UTC Range | Wavelengths / Filter Used | Airmass | Notes |
|---|---|---|---|---|---|---|
| Faulkes N. | Echeclus | Dec. 10 | 5:26 - 5:33 | r' filter | 1.221 - 1.223 | 6 x 60s |
| | Echeclus | Dec. 12 | 5:26 - 5:33 | r' filter | 1.892 - 1.190 | 6 x 60s |
| Faulkes S. | Echeclus | Dec. 13 | 11:00 - 11:07 | r' filter | 1.362 - 1.363 | 6 x 60s |
| | | | | | | |
| IRTF | Echeclus | Dec. 13 | 5:31 - 6:26 | 0.7 – 2.5 um | 1.16 – 1.06 | 4 x 200s* PA: 289 |
| | " | " | 9:29 - 10:08 | 0.7 – 2.5 um | 1.25 – 1.05 | 6 x 200s* PA: 66 |
| | SAO 93936 | " | 8:49 | 0.7 – 2.5 um | 1.010 | Solar Analog |
| | SAO 93028 | " | 5:21 – 10:15** | 0.7 – 2.5 um | 1.272 – 1.045** | Local Standard |

* The quoted number of exposures are the number of usable exposures. Those with poor or failed guiding were thrown out. Approximately ~1/2 of observations were lost for this reason.

** The local standard star observations were taken directly prior and after all science observations, not for the entire range listed.

**Figure 1: Spatial Profiles of 174P/Echeclus in NIR Spectra and Visible Imaging**

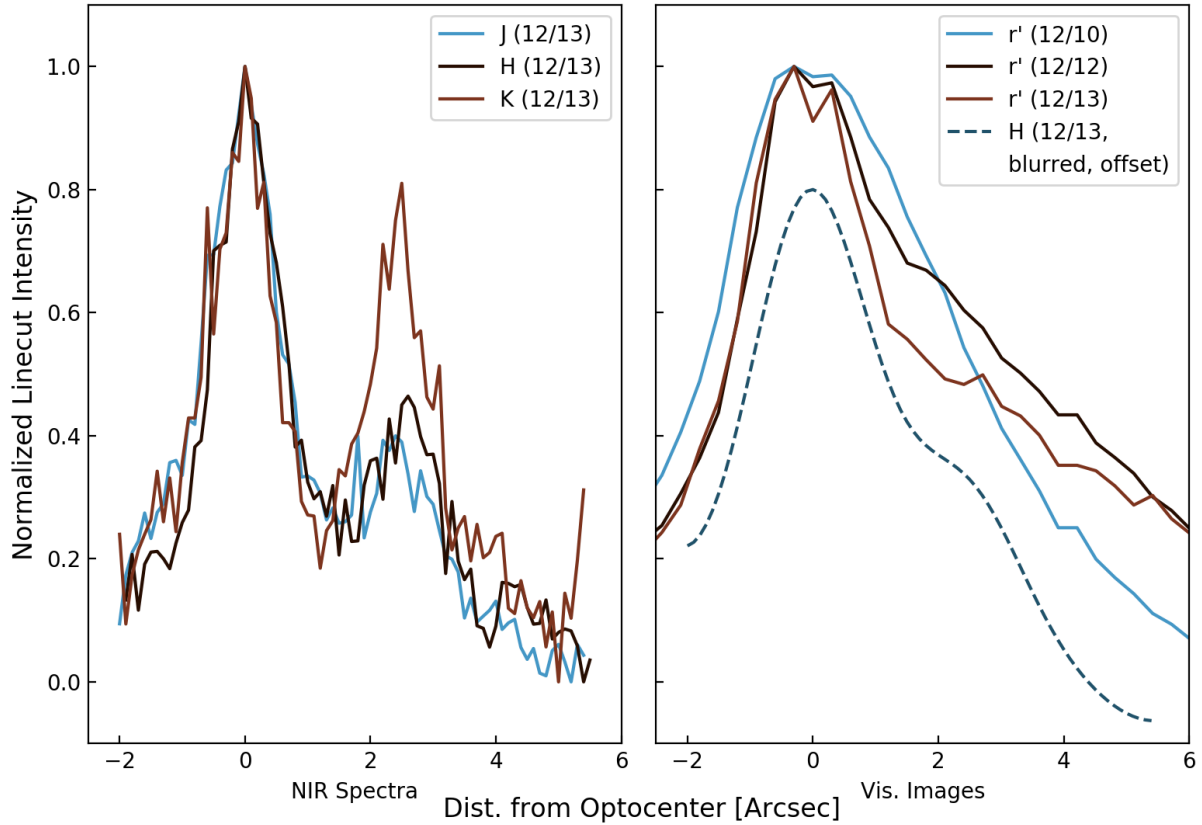

**Caption to Figure 1:** Left: The normalized along-slit spatial profiles extracted from a single 200s observation of Echeclus from wavelengths corresponding to the J, H, and K near-infrared filters. Two peaks are seen which are observed to track non-sidereally at the same rates and are ~2.45 - 2.6" apart at the time of observations. Each profile was divided by its peak value for cross-comparison of the shapes of the profiles, so the reader should not interpret that the J-Band data is the same brightness as the K-band data. Right: normalized spatial profiles along the same position angle as the near-infrared spectra shown at left extracted via a linecut technique from the visible images in the r' filter on the 10th, 12th, and 13th of December. The linecut technique uses a nearest-neighbor interpolation, so individual pixel-to-pixel changes are more susceptible to possible artifacts while overall trends are intact. A blurred version of the H-band profile is shown for comparison accounting for different seeing resolutions. The horizontal scale and limits of both plots are the same, and x-axis values increase along the direction of the position angle from the center of the coma. The asymmetry seen in the VIS data is seen to be compatible with the two-peaked structure in the NIR data when seeing is taken into account.

3.0 – Results and Analyses
3.1 – Imaging

The reduced images are shown cropped and centered on Echeclus in the top row of **Figure 2**. The bottom row of **Figure 2** shows the same images in the same order after the subtraction of an average radial profile to enhance the visibility of asymmetries in the coma. All images are normalized and centered on Echeclus's brightest pixel ("optocenter") and are shown with a log-scale color map, where brighter colors correspond to higher intensities. In the right-most image, the orientation of the NIR slit in the second block of observations (where the secondary peak was seen in the along-slit spatial profile) is added to the legend at top right. The coma is seen to be clearly asymmetric along the North-South direction even prior to image processing, and the difference becomes even more stark after radial profile subtraction. The north-south asymmetry is apparent in all of the images, though there are variations therein perhaps brought on by the continued release of new material, degradation of larger grains or macroscopic chunks, or other processes. Later imaging taken on Dec. 26th still shows the asymmetry clearly. This morphology, and the specifics of the asymmetry, bear great similarity to some outbursts of 29P detailed in Miles et al. (2016). Of note, in the Dec. 13 image closest in time to when our NIR spectroscopic observations are taken, the area of the coma brightest compared to the average radial profile is several degrees lower in position angle (e.g. further clockwise towards north) than the orientation of the slit was aligned to.

**Figure 2: Images of Echeclus's Expanding, Asymmetric Coma Post-Outburst**

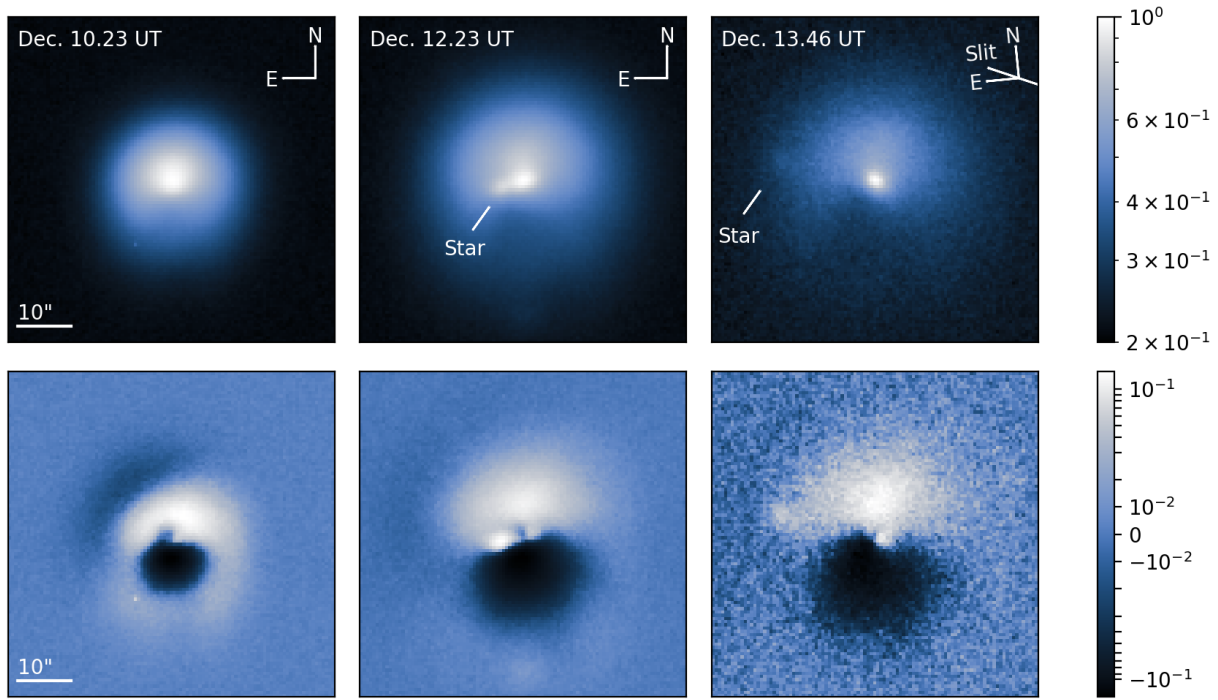

**Caption to Figure 2**: Top Row: Normalized and logarithmically scaled images centered on Echeclus's brightest pixel ("optocenter") between December 10th and December 13th UTC. Note the slight rotation on the Dec. 13th image, as noted by the rotated legend. The slit orientation of the second block of NIR spectral observations, in which a second peak was seen in the coma, is also noted on that legend. Bottom Row: The same images as the top row after the subtraction of an azimuthally averaged radial profile to enhance the visibility of asymmetric features in the coma, displayed on a logarithmic scale for both positive and negative values ('symlog'). For ease of cross comparison, all images are shown at the same spatial scale and normalized such that their brightest pixels are all white in the color maps utilized. We note that the star labeled in the December 13$^{th}$ is too far from the optocenter to affect our linecut analyses or spectra. Note that the color maps are not identical between the top and bottom row.

We also measured the diameter of the coma (left to right in the above images to avoid a bias due to the asymmetric coma) as a function of time to calculate its expansion rate and thus calculate an approximate time of outburst onset. Using a linecut technique, we find that on Dec. 10$^{th}$ at 10.23 UT, the coma is 15.0" +/- 0.7" in diameter, while on Dec. 12$^{th}$ at 10.23 UT, we find the coma to be 21.9" +/- 1.0" in diameter, or an expansion rate of 3.5 +/- 0.6 arcseconds per day. If instead of using a left-right cut, we analyzed the distance from the brightest pixel to where the coma reaches the sky background, we could retrieve an expansion rate of 3.3 +/- 0.7 arcseconds per day. (This allows us to probe the size of the coma Northwards in a way that does not simultaneously measure the dimmer Southern part of the coma.) Uncertainties quoted are from doing the same linecut or radial-distance technique at slightly different angles. Richard Miles, in a separate examination of the same images, found a similar expansion rate and an approximate time of onset as Dec. 7.05 +/- 0.20 (R. Miles, personal communication). James (2018) analyzed images of Echeclus submitted to the Comet Section of the British Astronomical Association (of which some of our images are included) and found an expansion rate of the coma diameter of 3.4"/day (~95 m/s at Echeclus's then geocentric distance), which our results agree with. They also found a relatively slow dimming of Echeclus's total brightness with time (~0.01 mag/day) after the outburst, which they suggest would be well explained by an essentially constant initial supply of dust so that as the coma expands the total area of dust reflecting sunlight remains the same. If this interpretation is correct, then the majority of the dust in the coma may have been released sufficiently early in the outburst to make continued ejection of material from the surface later on an unlikely culprit for the subtle variations in the coma morphology seen in our images.

The sparse in time images without obvious rotational features preclude creating a light curve or directly measuring the rotation period. However, in a similar set of observations of an outburst at 29P, Schambeau et al. (2017) used a Monte Carlo model of coma formation to conclude that the outburst duration was much shorter than the rotation period. However, more morphological structures were apparent after image processing in their data set than in ours, limiting both the applicability of their result to Echeclus and of a similar model to our data. In other words, the same kind of model applied to our data would have fewer constraints on the overall period, even if there are subtle features in our data set driven by Echeclus's rotation.

3.2 – Spectroscopy

The two reflectance spectra obtained with different extraction apertures are shown in **Figure 3**. The two spectra are red-sloped, gently curving, and lack absorption bands. The wider aperture spectrum is slightly bluer at shorter wavelengths.

**Figure 3: The Near-Infrared Reflectance Spectra of Echeclus Post-Outburst**

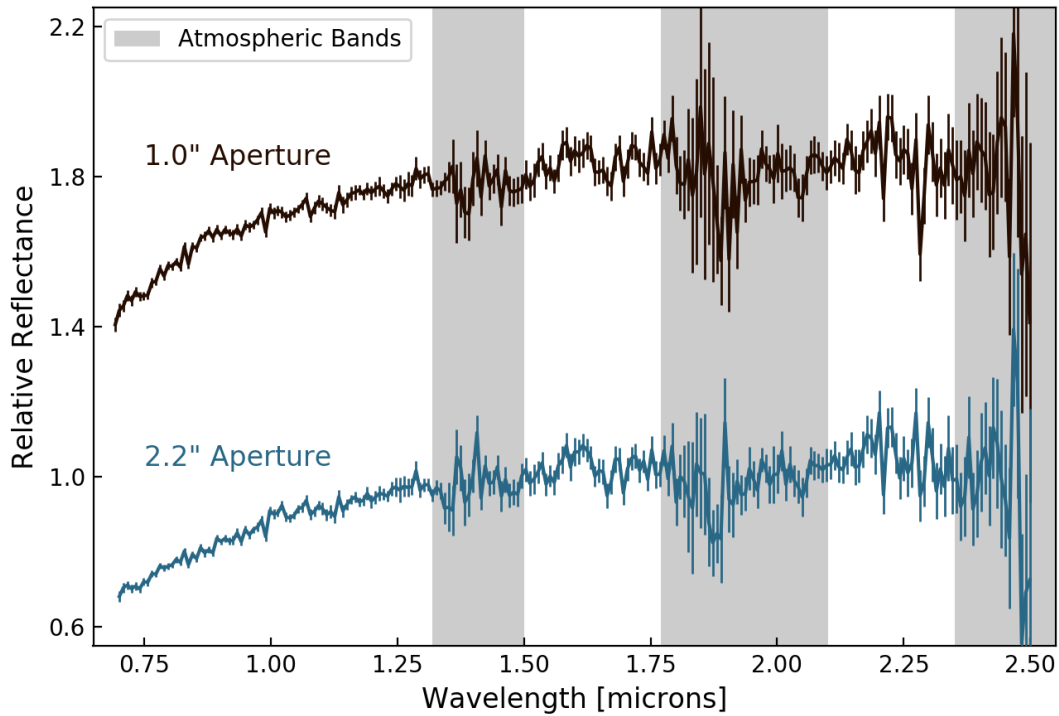

**Caption to Figure 3:** Reflectance spectra extracted using 1.0" and 2.2" apertures are shown with 1-sigma error bars, normalized at 1.5 microns and offset for clarity. Areas of low atmospheric transmission are shown as high-opacity gray horizontal lines behind the two spectra. The spectra are shown to both be red-sloped at wavelengths below 1.5 microns and nearly flat but still slightly red-sloped beyond that, which is consistent with previous observations of Echeclus (e.g. Guilbert et al., 2009, Secull et al., 2019). The wider aperture 2.2" spectrum is somewhat bluer (less red) than the narrower 1.0" spectrum. Quantitative measurements of the spectral slope are described in the text.

The two spectra presented in **Figure 3** are qualitatively very similar to previous spectroscopic observations of the nucleus of Echeclus (i.e. without known dust to pollute the spectrum), whereby a steeply red-sloped spectrum becomes more neutral (but still red) at longer wavelengths with a transition near ~1.0 - 1.2 microns (see below for quantification). There are no obvious absorption features above the noise in the spectra, though weak features could very easily be hidden. In general, our analysis is guided by healthy skepticism as our retrieved spectrum looks very much like previous observations of Echeclus's nucleus that had much longer integration times. This is discussed further in **Section 5**. The upper limits we could put on the strength of absorption features vary due to changing signal-to-noise as a function of wavelength, but are approximately a few percent at wavelengths shorter than the prominent atmospheric band (~1.3 microns) and could be as high as ~15% in K-band depending on location and where the continuum was calculated. The small up-and-down features observed in the data (see, for instance, the dip near ~1.67 microns) are highly sensitive to binning, and all such features were discounted. At the location of the water ice feature at 1.5 microns (not seen in our data), a feature would need to be 6%-10% less reflective than the surrounding continuum to be seen, while a feature at 2.0 microns (like the other band of water ice or $CO_2$) would need to be 10-15% deep compared to the local continuum even without considering skepticism from the atmospheric band at the same wavelength range. At wavelengths shorter than the slope break near ~1.0 - 1.2 microns, the 2.2" aperture spectrum, which should include more light reflected from dust in the coma, appears notably bluer (i.e. closer to neutral, but again, still red-sloped).

To compare our spectra to previous measurements, we calculated the spectral gradient using the same procedure as Seccull et al. (2019), whereby a line is fit to normalized data avoiding atmospheric bands and the slope is reported in percent per 0.1 microns. We note a slight difference between our technique and theirs: their data were fitted using a bootstrapping technique due to the spectral binning necessary to see the target, while we analyzed our unbinned data. The results of the following calculations are listed in **Table 4**. In the near-infrared, we normalized our two versions of Echeclus's spectrum at 1.6 microns and then used data between 1.25 and 1.7 microns, excluding that in the atmospheric band between 1.3 and 1.5 microns. This is listed as "NIR" in **Table 4**. In the visible (essentially the R and I photometric bands), Seccull et al. (2019) normalized their spectra at 0.658 microns, which is outside of our wavelength range by just ~0.04 microns, so we adopt two methods to calculate our spectral gradients in this wavelength range. The first attempts to replicate their normalization by extrapolating an initial fit to the data down to 0.658 microns to normalize it there and then fitting all of our data shortwards of 0.8 microns using that normalization. This is listed as "VIS" (Method 1) in **Table 4**. However, this method uses only ~0.105 microns of our spectrum and thus could fail to capture the difference between our two derived spectra. The second method is to fit the spectral gradients with a normalization at 1.6 microns (the same as our NIR gradients) and fit all data shortward of 0.9 microns. This is listed as "VIS" (Method 2) in **Table 4**.

**Table 4: Calculated Spectral Slopes**

|  | 1.0" Aperture | 2.2" Aperture | Secull et al., 2019. |
| --- | --- | --- | --- |
| "NIR" | S' = 1.34 +/- 0.11 | S' = 1.32 +/- 0.10 | S' = 1.26 - 1.29 |
| "VIS" (Method 1) | S' = 10.96 +/- 0.50 | S' = 8.18 +/- 0.39 | S' = 11.39 – 12.12 |
| "VIS" (Method 2) | S' = 10.76 +/- 0.16 | S' = 6.57 +/- 0.13 | N/A |

In general, both our 1.0" and 2.2" aperture spectra are fully consistent with previous observations of the reflectance spectrum of Echeclus's nucleus in the absence of (known) dust beyond ~1.0 microns. However, at wavelengths shorter than ~1.0 micron, the two spectra we retrieved with different apertures are statistically very different using either of the fitting methods described above. The 1.0" aperture spectrum is statistically consistent with previous observations of the bare nucleus in the optical, while the 2.2" aperture spectrum is significantly bluer than both its 1.0" counterpart and previous observations of Echeclus's surface regardless of fitting method employed. The wider aperture should include more reflected light from dust in the coma, which has previously been observed to be blue (e.g. Bauer et al., 2008, Seccull et al., 2019), which we discuss further in **Section 5**. We note here, and in **Section 5**, that the decreasing importance of dust at longer wavelengths helps to explain the increasing sharpness of both peaks seen in the NIR along-slit spatial profile and the lack of secondary peak seen in the visible data along the most-enhanced part of the coma.

4.0 – *Echeclus's Recent Orbital History*

Echeclus's strange activity patterns – medium to very large outbursts throughout its orbit with little notable activity in between, fragmentation or debris ejection event(s?) – must be related in some way to its recent history and evolution. While Echeclus's orbital history has been studied in the past (Gladman et al., 2008), the focus was on dynamically classifying the object as opposed to identifying any recent trends in the evolution of its orbit that might put its modern activity in context. To help frame our discussion of Echeclus's behavior and determine if it is representative of the Centaur population at large, we integrated Echeclus's orbital motion and that of 100 orbital 'clones' backwards in time using the IAS15 integrator within the orbital integration code Rebound (Rein and Spiegel 2015). The nominal orbit was taken from JPL Horizons, with the clone orbits spanning the orbital uncertainty by drawing from the covariance matrix for that orbit available on the JPL Small Body Database Browser. We integrated the nominal and clone orbits forward 1,000 years and backwards 10,000 years as massless test particles to both gain quantitative inferences about Echeclus's thermal environment, as well as identify any notable close encounters with the giant planets. We included the effects of the Sun and the planets Mercury through Neptune with a timestep of 0.001 years, or approximately 8.75 hours. We note that the IAS15 integrator uses an adaptive timestep to properly resolve close gravitational encounters between objects, which is critical for the (by definition) often gravitationally perturbed Centaur population. The evolution of Echeclus's semimajor axis ('a') and perihelion distance ('q') from these simulations between 1000 BCE and 3000 CE are shown in **Figure 4**.

**Figure 4: Recent Dynamical History of 174P/Echeclus**

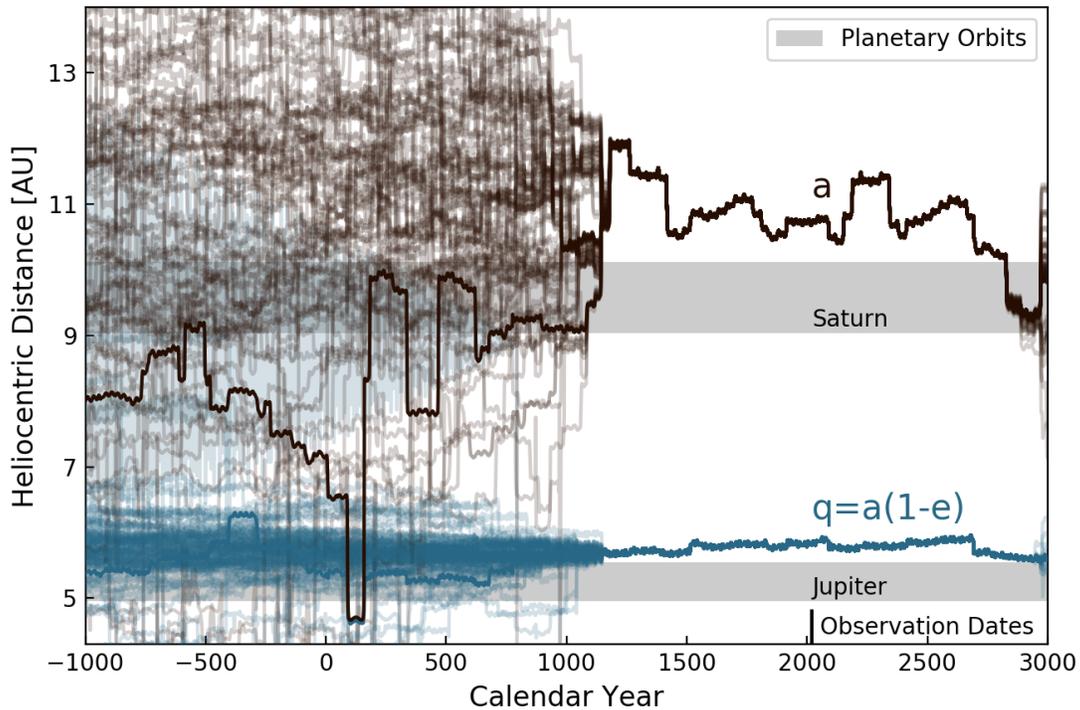

**Caption to Figure 4.** The evolution of Echeclus's orbital parameters (semimajor axis 'a' and perihelion distance 'q') compared against 100 "clones" of its orbit within several thousand years of the present day. The clones are plotted with a lower opacity in the same color as the nominal orbit. The orbits of Jupiter and Saturn are shown as shaded boxes for context. The orbits of the clones diverge from the nominal orbit around the year 1144, before which the orbit can only be described statistically. Echeclus's orbit since then has had a stable but slightly varying perihelion distance near 5.6-5.9 AU and a semimajor axis varying between 10.3 and 12.0 AU. Unlike many active Centaurs (Fernandez et al., 2018) Echeclus's orbit shows no evidence for a recent drop in perihelion distance, which could indicate a change in thermal environment.

The nominal orbit and the orbits of the clones remain very similar between approximately ~1144 CE (a close approach with Jupiter at 0.90 AU) and ~2970 CE (a further approach with Jupiter at 1.98 AU), and beyond that are only statistically known. Echeclus's perihelion distance of 5.6-5.9 AU has remained fairly stable for the past millennium, while the semimajor axis varies between 10.1 and 12.0 AU. Within the time frame where the orbit is well-determined, there are many gravitational encounters with Jupiter and Saturn but none are strong enough to change the orbit drastically. The closest encounter with Jupiter is 1.559 AU (~4.5 Jupiter Hill Radii) in the year 1416, while with Saturn it was 1.02 AU (~2.5 Saturn Hill Radii) in the year 1513. Within the well-determined period, there have been small trends towards increasing perihelion distance and decreasing semimajor axis - e.g. increasing eccentricity with time.

Fernandez et al. (2018) show that for many currently active Centaurs, there has been a significant decrease in perihelion distance in the recent few hundred to a few thousand years, which they interpret as evidence for increased temperatures on the surface and in the sub-surface, allowing for more devolatilization and advancing crystallization of water ice to drive the activity (should the amorphous water ice actually be present). This was also noted and discussed for 29P in Sarid et al., (2019). This explanation certainly is certainly plausible with our limited understanding of Centaur / comet properties, but it fails to explain activity of objects in orbits that have not undergone recent changes, such as 95P/Chiron (Fernandez et al., 2018) or Echeclus. Considering that the temperature at the subsolar point of a planetary body at any point in its orbit should be proportional to the inverse of the square root of the heliocentric distance, these small changes in the orbit of Echeclus do not drastically change the thermal environment it is in. (The peak temperature changes approximately ~3% between a perihelion at 5.6 AU and 5.9 AU, and the average temperature changes ~8% between semimajor axes of 10.1 AU and 12 AU for reference.) Echeclus, like Chiron, cannot have its activity explained simply through large orbital changes inducing higher temperatures, as those changes have not occurred in the recent past. However, any small change in orbit should change the thermal state of the nucleus, if only slightly. A different distribution of temperatures combined with a changing topography and/or changing rotational pole orientation due to comet-like outgassing and mass-wasting could open up new parts of the nucleus to continued sublimation until all of the volatiles within several thermal skin depths of the surface have been released. In this picture, a nucleus re-equilibrating would be expected to have higher levels of activity at perihelion when temperatures are overall the highest (seasonality notwithstanding). However, we know that Echeclus's modern activity is both episodically spread along its entire orbit observed thus far and has very intense outbursts, which does not easily fit into a picture of a perturbed nucleus moving towards thermal equilibrium inside its orbit.

5.0 – Discussion

In this section, we detail our interpretation on the properties of the dust in the coma of Echeclus and the properties of the secondary source we see inside of it. We close with a brief discussion of how Echeclus's activity and properties reflect on its place within the Centaur population.

5.1 – Dust Properties

The similarity in coma morphology between our visible wavelength images and near-infrared 2D spectra (see **Figure 1**) when seeing is accounted for suggests that they're "seeing" the same dust. While a narrow aperture extraction retrieved a reflectance spectrum consistent with Echeclus's nucleus, the reflectance spectrum became bluer upon extraction with a wider aperture, suggesting that the dust was likely bluer than the nucleus. Echeclus's dust has previously been observed to be blue spectrally in Seccull et al. (2019), which our measurements are consistent with. A spectrally blue coma is a somewhat rare property among comets but was also seen in the larger Centaur 95P/Chiron (West, 1991). The two common explanations for this behavior is that the dust is simply blue in color or that the coma is dominated by smaller grains that are less efficient at scattering longer wavelengths of light. Secull et al. (2019) argued for blue grains, as opposed to small ones, for two primary reasons. First, and perhaps most importantly, Jewitt (2015) argues that cometary comae are usually dominated in effective cross section by the largest grains. As Echeclus in its larger 2005 outburst appeared to have "normal" cometary activity (Bauer et al., 2008), this should likely apply to Echeclus's coma as well. Second, the smaller grains should be accelerated to higher speeds than the larger ones, and Seccull et al. (2019) performed their observations several weeks after the 2016 outburst of Echeclus. If the smallest grains are moving the fastest, then our measurements of the coma expansion (~3-4"/day) rate place them far from the spectral extraction aperture, and thus unlikely to contribute greatly. Grains still inside our 0.8" x 1.0" (2.2") aperture ~6 days after the presumed onset of outburst should only be moving ~5 m/s or less, compared to the ~95 m/s overall coma expansion speed. If we assume that the velocity of an individual dust grain is proportional to its radius raised to the -1/2 power (energy conservation), then the particles still inside our aperture are likely more than 300 times larger in radius than the smallest grains available. Additionally, there are many processes by which cometary or comet-like materials can be made to appear blue visually, from the inclusion of carbon-bearing materials that sometimes appear blue (Cloutis et al., 1994, Clark et al., 2010) to size sorting (Clark et al., 2010, Hiroi et al., 2010, Cloutis et al., 2013) or phase angle (Cloutis et al., 2011) effects. The break-down and dehydrogenation of complex carbon-chain molecules can also make them bluer (Fomenkova et al., 1994, Cloutis et al., 1994).

Thermal infrared observations presented in Bauer et al. (2008) showed that not only was Echeclus's coma dominated by large dust grains, but also that the morphology was quite similar at visible and thermal infrared wavelengths. The similarities in strength (>4 visual magnitude increase in brightness), the existence of a secondary condensation, and the similarity of morphologies in our 2017 visible and near-infrared observations suggest at the very least qualitative similarities between the 2005 outburst and the 2017 outburst described in this paper, and perhaps similarities between their causes as well. In light of that, we argue that the presence of large dust grains that are blue in color is the simplest explanation for our observations as well. The origin of the blue color of these primarily large grains remains open, however. Future time-domain near-infrared imaging could look for color gradients within the coma as it evolves, which might better diagnose whether or not the grains are expelled from the surface already 'blue' or become blue as they are size-sorted and degrade in the coma.

As mentioned in **Section 4**, previous observations of Echeclus's nucleus (Guillbert et al, 2009, Seccull et al., 2019) noted no obvious ice absorption features on the surface. Our data also has no obvious ice features, but it is plausible that whatever material was excavated and ejected from the surface to form the dust coma might either contain some new ices or leave ices newly exposed on the nucleus. As previous observations (especially those of Seccull et al., 2019) had longer effective exposure times, and our observations in the near-infrared appear to be consistent with reflected light from the nucleus, we interpret this that the nucleus did not have any massive changes in surface composition and the dust is not incredibly icy, neither of which are surprising.

5.2 – The Secondary Source

As mentioned initially in **Section 2** and shown in **Figure 1**, the spatial profiles of our near-infrared spectra from the second block of observations show two separate peaks in the coma. The secondary peak is consistently placed 2.45" – 2.60" from the primary peak throughout multiple consecutive exposures, suggesting that it is moving at the same non-sidereal tracking rate as Echeclus itself. The secondary peak is in the brighter part of Echeclus's coma and is hinted at very strongly in the line cut profiles extracted along the same position angle as the second block's slit orientation.

The most tantalizing explanation is that Echeclus has again released a large fragment or condensed mass of material, most likely corresponding to the start of the outburst, analogous to the 2005 event most commonly interpreted as a large-scale fragmentation event. If we assume that the cause of the second peak was ejected from the surface of Echeclus at our nominal time of outburst onset, the 2.45-2.6" separation at a geocentric distance of 6.53 AU results in estimated speeds of 21 – 23 m/s, which suggests it is moving hyperbolically relative to the primary nucleus, another feature shared with the 2005 event (Weissman et al., 2006b). James (2018) estimated the overall coma expansion rate (essentially the dust shell expansion rate) to be ~95 m/s, much faster than our estimated speed above. If we assume the smallest dust grains are accelerated to the fastest speeds, then we might then assume the material that makes up our second peak is then larger and decoupled from the gas quicker.

The theoretical prospect of ejecting a large macroscopic fragment at such a speed was and remains fundamentally unsolved. However, ejecting a large amount of debris even up to several meters in size is more than possible if activity is highly localized. In the absence of unambiguous evidence for a large 2005-like macroscopic fragment, the latter scenario seems like the best explanation for our observations. In that framework, the secondary peak is likely a collection of material up to a few meters in size ejected in approximately the same direction from the surface of Echeclus. While there are only weak constraints on the size of the 2005 fragment, Echeclus seems to be able to eject material to and above its own escape velocity from dust a few microns in size through boulders up to even larger full fragments. More modeling work is critical to understanding these processes and to understanding the Centaurs in general.

5.3 – Echeclus among the Centaurs

Is Echeclus 'typical' among the Centaurs? It has a reflectance spectrum typical of many active Centaurs (Peixinho et al., 2003), but its recent orbit has changed far less than many other active Centaurs (Fernandez et al., 2018). It experiences large outbursts like many of the best-studied Centaurs (29P, Chiron) and long period comets at similar distances, but is apparently dormant or only minimally active in between, unlike 29P. As we have only studied a handful of Centaurs to any great depth, the Copernican interpretation would be that Echeclus is at least somewhat typical, as are the other best-known Centaurs, in the absence of other information. If that is the case, then there is reason to think that many other active Centaurs are also releasing fragments and large clumps of debris as well. If Echeclus has had two of these explosive debris ejections in ~12 years, and the behavior is shared by some fraction of the active population, then it is possible that several of these fragmentation/debris ejections are being missed *each year* on other Centaurs. The active Centaurs are an incredibly diverse group, from barely active to the most active objects in the solar system, from constantly active to intermittent at most, and they display many behaviors not seen in the inner solar system comets and not yet well understood. More observational resources could be instrumental in both understanding these behaviors better as well as finding new processes not yet observed.

6.0 – Summary

Understanding the Jupiter-Family-Comet/Centaur/Kuiper-Belt-Object relationship is critical to understanding the history of any of the three groups. Among the Centaurs, some fraction show cometary activity, some of which is unlike the kind of activity seen among inner solar system comets. Jupiter Family Comets likely "turn on" for the first time as Centaurs and understanding how and why is an area of much ongoing research. We present new visible-wavelength imaging and near-infrared spectroscopy of the active Centaur 174P/Echeclus in the first week after its December, 2017 outburst. The coma is observed to be highly asymmetric in the North-South direction, expanding 3.3 – 3.5 arcseconds per day, and a likely time of onset of Dec. 7.05 +/- 0.20. The near-infrared 2D spectra show two peaks, a shape hinted at strongly in the visible images obtained with worse seeing. The separation between the peaks suggests an approximate speed of ~21 – 23 m/s, significantly slower than the overall coma expansion speed but higher than the escape velocity of the nucleus. The retrieved NIR reflectance spectrum of Echeclus is consistent with the bare nucleus if a small (1.0") extraction aperture is used but becomes significantly bluer if a larger (2.2") extraction aperture is used. The similarity of morphologies between the NIR and VIS data, as well as the color of the dust (bluer than the nucleus), is broadly consistent with the ideas of Bauer et al. (2008) and Seccull et al. (2019) that Echeclus has a large-grained and blue coma. We perform a high-resolution orbital integration, which reveals that unlike most of the active Centaurs it has had a relatively stable orbit for the past ~900 years (Fernandez et al., 2018); this rules out a recent thermal change due to orbital evolution as the explanation for Echeclus's outbursts. We argue that the second peak seen in our data is composed of large debris (bigger than dust, smaller than the 2005 fragment) ejected at the onset of the outburst. More theoretical work is critical in understanding these processes, and more observational resources could help to determine how common they are among the active Centaurs. If Echeclus is typical, then there may be many kinds of behaviors and activity patterns among the Centaurs that have not been noted yet.

Acknowledgment

This research work was supported by NASA Near-Earth Object Observations Grant NNX17AJ19G (PI: Reddy). We thank the IRTF TAC for awarding time to this project, and to the IRTF TOs and MKSS staff for their support. The authors wish to recognize and acknowledge the very significant cultural role and reverence that the summit of Mauna Kea has always had within the indigenous Hawaiian community. We are most fortunate to have the opportunity to conduct observations from this mountain. Perceptually uniform color maps are used in this study to prevent visual distortion of the data and to better serve those with color vision deficiency (Crameri 2018a,b).